\documentclass{Interspeech2024}
\usepackage{multirow}
\usepackage{graphicx}
\usepackage{caption}
\usepackage{amsmath,graphicx,hyperref}

\interspeechcameraready

\title{The TCG CREST –- RKMVERI Submission for the NCIIPC Startup India AI Grand Challenge}

\name[affiliation={1,2}]{Nikhil}{Raghav}
\name[affiliation={2}]{Arnab}{Banerjee}
\name[affiliation={2}]{Janojit}{Chakraborty}
\name[affiliation={1}]{Avisek}{Gupta}
\name[affiliation={2}]{Swami}{Punyeshwarananda}
\name[affiliation={1}]{Md}{Sahidullah}

\address{
  $^1$Institute for Advancing Intelligence, TCG CREST, Kolkata-700 091, India\\
  $^2$Department of Computer Science, RKMVERI, Howrah-711 202, India}
\email{nikhil.raghav.92@tcgcrest.org}


\begin{document}

\maketitle

\begin{abstract} 
In this report, we summarize the integrated multilingual audio processing pipeline developed by our team for the inaugural NCIIPC Startup India AI GRAND CHALLENGE, addressing Problem Statement 06: Language-Agnostic Speaker Identification and Diarisation, and subsequent Transcription and Translation System. Our primary focus was on advancing speaker diarization, a critical component for multilingual and code-mixed scenarios. The main intent of this work was to study the real-world applicability of our in-house speaker diarization (SD) systems. To this end, we investigated a robust voice activity detection (VAD) technique and fine-tuned speaker embedding models for improved speaker identification in low-resource settings. We leveraged our own recently proposed multi-kernel consensus spectral clustering framework, which substantially improved the diarization performance across all recordings in the training corpus provided by the organizers. Complementary modules for speaker and language identification, automatic speech recognition (ASR), and neural machine translation were integrated in the pipeline. Post-processing refinements further improved system robustness. 
\end{abstract}

\section{Introduction}
To address emerging challenges in the context of the national security landscape in India. Innovative technological solutions are paramount in an increasingly intricate digital environment. Recognizing this imperative, the National Critical Information Infrastructure Protection Centre (NCIIPC) launched the \emph{Startup India AI Grand Challenge}\footnote{\url{https://ai-grand-challenge.in/}}. This initiative lays a comprehensive foundation to harness artificial intelligence and machine learning capabilities from India's Startup ecosystem, or student groups who are willing to be a startup, to strengthen the national security infrastructure across multiple critical domains, ranging from Cybersecurity \& Forensics to Maritime and Signal Processing domains. 

\begin{figure}[t]
  \centering
  \includegraphics[width=1.05\columnwidth]{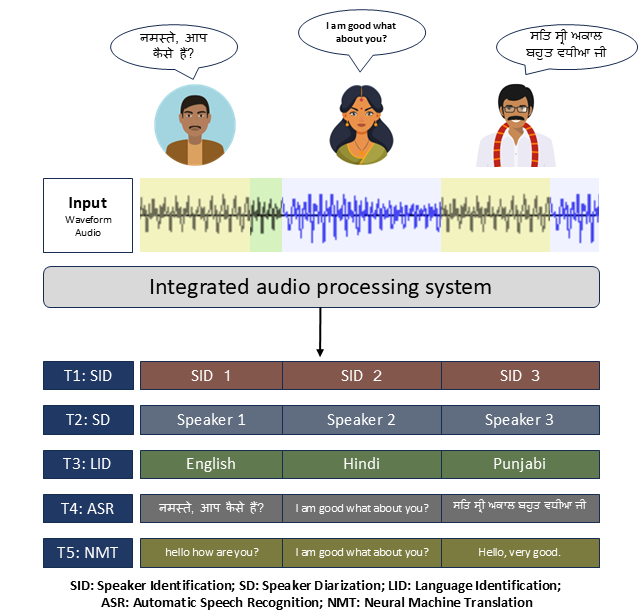}
  \caption{The five tasks in the problem statement PS-06.}
  \label{fig:pipeline1}
\end{figure}

Our focus is on the problem statement \textbf{PS-06} concerning real-life research applications in conversational AI, including Language Agnostic Speaker Identification \& Diarisation; and subsequent Transcription \& Translation System, as shown in Figure~\ref{fig:pipeline1}. Since in Phase 1 the concerning datasets include multi-lingual and code-switching and code-mixing in the Hindi, English, and Punjabi languages, respectively. Our objective was to develop an integrated audio processing pipeline that operates on such data and provides solutions to all five problems outlined in the problem statement. However, our main intent was to assess the real-world applicability of our speaker diarization (SD) system under low resource scenario in Indian context.

\section{Overview of the Problem}
The problem statement, PS-06, aims to develop an integrated audio processing system that transforms spoken audio into structured, multilingual, and textual insights. 
Given an input audio file, the system should be able to perform the following tasks in offline mode: 
\begin{enumerate}
    \item \textbf{Speaker Identification (SID):} Match each speaker segment to a known identity when enrollment data is available~\cite{SID,SID-2}.
    \item \textbf{Speaker Diarization (SD):} Segment the audio by identifying the boundaries between different speakers~\cite{PARK2022101317}.
    \item \textbf{Language Identification (LID):} Detect the language spoken in each segment, supporting multilingual and code-switched audio~\cite{language_recognition, dey2024towards}.
    \item \textbf{Automatic Speech Recognition (ASR):}  Convert each speaker’s speech into accurate text in spoken script~\cite{ASR}. 
    \item \textbf{Neural Machine Translation (NMT):} Translate the transcribed text into the English language, preserving speaker-level segmentation~\cite{stahlberg2020neural}.
\end{enumerate}

The solution should be able to handle audio files with the following dependencies: 
\begin{enumerate}[label=(\alph*)]
    \item Sample rate: between 8 kHz to 48 kHz
    \item Bit depth: between 8 to 32 bits
    \item File types: should cover at least \texttt{wav}, \texttt{mp3}, \texttt{ogg}, \texttt{flac}
    \item Signal to noise ratio (SNR): 5 dB or higher (i.e., a higher value indicates better quality)
\end{enumerate}

We, as the participants, were advised to use our own or open-source datasets (e.g., AI Kosh\footnote{\url{https://aikosh.indiaai.gov.in/}}) for training the model. However, the datasets (audio files) should conform to the specifications mentioned above. In addition, they should include natural overlaps, code-switching, reverberations, and noise to ensure robustness and real-world diversity.

\section{Our Approach}
The tasks of SID and SD operate independently, whereas the remaining tasks, LID, ASR, and NMT, operate sequentially. In the last three tasks, the output of the former serves as an input to the latter system. The overall structure of the pipeline is as shown in Figure \ref{fig:pipeline}. 

\begin{figure}[t]
  \centering
  \includegraphics[width=0.9\columnwidth]{Pipeline.drawio.png}
  \caption{Our approach for building an integrated audio processing pipeline. The first block of VAD serves as an input to SID, LID, and SD, respectively. Further, the output of the LID serves as an input to the ASR, and likewise for the NMT module.}
  \label{fig:pipeline}
\end{figure}

\section{Descriptions of the Modules}

\subsection{Speech Activity Detection}
For speech activity detection (SAD), we used the pre-trained neural Silero-VAD~\cite{Silero-VAD} model\footnote{\url{https://github.com/snakers4/silero-vad}}. For each recording in the test dataset, the corresponding start and end timestamps are exported to a CSV file.

\subsection{Speaker Diarization}
Our speaker diarization system employs a fully unsupervised clustering method that we recently investigated~\cite{Raghav2026MKSGCSC}, which outperformed conventional spectral clustering approaches. We adopted scripts from the SpeechBrain toolkit for the diarization experiments. The main steps of the diarization pipeline are as follows:

\begin{itemize}
    \item \textbf{Converting SAD labels:} The SAD labels generated using SileroVAD are converted into a pseudo RTTM file to fit the SpeechBrain~\cite{speechbrain_v1, speechbrain} pipeline~\footnote{\url{https://github.com/speechbrain/speechbrain/tree/develop/recipes/AMI}}. The RTTM file contains the start timestamps and durations of the detected speech segments, which are further segmented into chunks of 1.00~s with an overlap of 0.50~s.

    \item \textbf{Speaker embedding extraction:} We use ECAPA-TDNN~\cite{desplanques20_interspeech} embeddings extracted with a pre-trained model trained on the VoxCeleb dataset\footnote{\url{https://huggingface.co/speechbrain/spkrec-ecapa-voxceleb}}. The audio sampling rate is converted to 16,000~Hz when required.

    \item \textbf{Unsupervised clustering:} We apply the fully unsupervised clustering method~\cite{Raghav2026MKSGCSC} an extension of~\cite{SC-pNA} using exponential and arccosine1 kernels with 15 nearest neighbours.
\end{itemize}

\subsection{Speaker Identification}
Our speaker identification (SID) system employs a pre-trained ECAPA-TDNN speaker verification model trained on Voxceleb1 and Voxceleb2 datasets. The main steps of the speaker identification pipeline are as follows: 
\begin{itemize}
    \item \textbf{VAD Segmentation:} Speech segments are extracted using Silero-VAD~\cite{Silero-VAD}.
    
    \item \textbf{Speaker enrollment:} For each enrollment speaker, ECAPA-TDNN speaker embeddings are extracted. Further, the mean of the embedding forms the speaker centroid. 
    
    \item \textbf{Processing test data:} Each audio from the test data is split into 3.00~s chunks based on the VAD information.
    
    \item \textbf{Scoring:} Each chunk of embedding is compared to the centroids using cosine similarity. 
    
    \item \textbf{Post processing:} To stabilize speaker predictions and reduce abrupt label changes across adjacent segments, median filtering is applied over the sequence of predicted speaker labels. The filter window size (\textit{smooth\_k}) is empirically chosen to balance temporal smoothness and responsiveness to genuine speaker changes.
\end{itemize}

\subsection{Language Identification}
Our language identification (LID) system employs the pre-trained VoxLingua107\footnote{\url{https://huggingface.co/speechbrain/lang-id-voxlingua107-ecapa}} ECAPA-TDNN Spoken Language Identification Model. The main steps of the language identification pipeline are as follows: 
\begin{itemize}
    \item \textbf{Data Preparation:} To create a CSV file with the following columns in order: AudioFile, Language, Start, End.
    
    \item \textbf{Feature Extraction:} For each 3.00~s chunk, an embedding is extracted using the ECAPA-TDNN model.
    
    \item \textbf{Training:} Trained a Logistic regression classifier on the ECAPA-TDNN embeddings using the labeled Mock data consisting of 20 audio files, shared by the organizers.  
    
    \item \textbf{Inference:} For each 3.00~s chunk of audio segment from the test data set, a language label is predicted using the trained classifier with a confidence score. 
\end{itemize}

\subsection{Automatic Speech Recognition}
Our automatic speech recognition (ASR) system employs the following two transcribers: Whisper-Small.en for English, and ai4bharat IndicWhisper-hi for both Hindi and Punjabi languages. The main steps involved in the ASR pipeline are as follows: 
\begin{itemize}
    \item \textbf{Segmentation: } Extracted 3.00~s chunks of segments from the input audio to the system.
    
    \item \textbf{Model selection: } The language ID labels curated from the LID inference for each segment act as a guide for the model selection. 
    
    \item \textbf{Transcription: } The model processes each chunk, extracts features, generates predicted IDs, and decodes the text, in the source language. 
    
    \item \textbf{Post-processing: } For the Punjabi text, the model transliterate from Devnagari to Gurmukhi. 
    
\end{itemize}

\subsection{Speech Translation}
Our neural machine translation (NMT) system comprises the following two models. The IndicTrans2 is the primary model responsible for translation from Hindi to English and Opus-MT is for Punjabi to English translation. The main steps involved in the NMT pipeline are as follows:
\begin{itemize}
    \item \textbf{Read the ASR CSV:} For each row, the following information, AudioFile, StartTS, EndTS, Transcript, is read.
    \item \textbf{Detect the language:} To detect the language per segment among hi (Hindi), pa (Punjabi), and en (English).
    \item \textbf{Split Code-switched spans:} The mixed-language segments are further split to be compatible with the model.  
    \item \textbf{Load the translator:} Loads the translator models one per language.
    \item \textbf{Translate spans:} For the English segments, no model is selected, and the transcription passes through. Beam search is used with a beam size of 5 to handle the language tag and to guide the model selection. Translation is carried out accordingly from the source language to English.
    \item \textbf{Finalize and merge:} The translation is finalized and merged for all the audio recordings in the test data.
\end{itemize}

\begin{table}[t]
\renewcommand{\arraystretch}{1.2}
\centering
\caption{Recording-level audio properties across the dataset.}
\setlength{\tabcolsep}{3pt} 
\footnotesize 
\resizebox{\columnwidth}{!}{%
\begin{tabular}{|l|c|c|c|c|c|}
\hline
\textbf{File} & \textbf{Type} & \textbf{Sampling Rate (Hz)} & \textbf{Bit Rate (kbps)} & \textbf{Duration (s)} & \textbf{SNR (dB)} \\
\hline
ID1 & wav  & 16,000 & 256 & 05:02 & 7.38 \\
ID2 & wav  & 22,050 & 352 & 07:06 & 12.21 \\
ID3 & wav  & 32,000 & 512 & 03:17 & 13.92 \\
ID4 & wav  & 44,100 & 705 & 05:06 & 22.72 \\
ID5 & wav  & 88,200 & 2822 & 02:51 & 19.99 \\
ID6 & wav  & 96,000 & 3072 & 04:13 & 15.74 \\
ID7 & wav  & 176,400 & 5644 & 06:44 & 18.78 \\
ID8 & wav  & 192,000 & 3072 & 06:01 & 15.05 \\
ID9 & wav  & 352,800 & 5644 & 04:52 & 11.83 \\
ID10 & wav  & 16,000 & 256 & 04:15 & 18.55 \\
ID11 & mp3  & 8,000 & 29 & 06:43 & 9.42 \\
ID12 & mp3  & 12,000 & 41 & 04:39 & 12.70 \\
ID13 & ogg  & 11,025 & 176 & 06:34 & 9.21 \\
ID14 & wav  & 8,000 & 128 & 03:44 & 25.98 \\
ID15 & wav  & 8,000 & 64 & 04:13 & 10.83 \\
ID16 & ogg  & 32,000 & 512 & 06:26 & 12.33 \\
ID17 & flac & 44,100 & 653 & 06:10 & 15.81 \\
ID18 & wav  & 8,000 & 128 & 08:58 & 18.69 \\
ID19 & wav  & 16,000 & 256 & 05:55 & 14.89 \\
ID20 & wav  & 32,000 & 512 & 05:34 & 17.87 \\
ID21 & wav  & 48,000 & 768 & 08:07 & 20.62 \\
\hline
\end{tabular}%
}
\label{tab:recording_properties}
\end{table}

\begin{table*}[t]
\renewcommand{\arraystretch}{1.2}
\centering
\caption{File-level performance across tasks: VAD (Accuracy), SID (IER), LID (DER), ASR (WER), and NMT (BLEU). Lower values indicate better performance for IER, DER, and WER, whereas higher values indicate better performance for Accuracy and BLEU.}
\setlength{\tabcolsep}{4pt} 
\resizebox{\textwidth}{!}{%
\begin{tabular}{|c|c||ccc||c|c|c|c|}
\hline
\textbf{File} & \textbf{VAD Acc. (\%)} & \textbf{DER (CSC)} & \textbf{DER (MK)} & \textbf{RI (\%)} &
\textbf{SID IER (\%)} & \textbf{LID DER (\%)} &
\textbf{ASR WER (\%)} & \textbf{NMT BLEU} \\
\hline
ID1 & 93.507 & 72.93 & 17.34 & 76.2 & 100 & 177.77 & 87.97 & 0.205 \\
ID2 & 89.162 & 14.45 & 14.30 &  1.0 & 100 & 24.98 & 38.04 & 0.395 \\
ID3 & 90.412 & 20.28 & 11.35 & 44.0 & 100 & 39.18 & 28.05 & 0.505 \\
ID4 & 88.946 & 13.65 & 13.82 & -1.2 & 100 & 31.00 & 43.73 & 0.388 \\
ID5 & 86.236 & 71.01 & 19.33 & 72.8 & 100 & 22.03 & 30.10 & 0.466 \\
ID6 & 92.084 & 11.21 & 10.46 &  6.7 & 100 & 21.10 & 51.16 & 0.333 \\
ID7 & 96.304 &  8.34 &  8.60 & -3.1 & 100 & 4.51 & 49.43 & 0.311 \\
ID8 & 92.243 &  8.59 &  8.59 &  0.0 & 100 & 5.62 & 88.45 & 0.259 \\
ID9 & 86.783 & 30.73 & 29.63 &  3.6 & 100 & 8.31 & 55.28 & 0.388 \\
ID10 & 60.494 & 52.29 & 48.41 &  7.4 & 100 & 19.23 & 73.18 & 0.101 \\
ID11 & 81.281 & 43.56 & 44.11 & -1.3 & 100 & 4.43 & 69.33 & 0.037 \\
ID12 & 93.252 & 22.42 & 18.37 & 18.1 & 100 & 26.97 & 51.95 & 0.138 \\
ID13 & 91.980 &  9.76 &  9.76 &  0.0 & 108 & 20.23 & 38.50 & 0.175 \\
ID14 & 94.327 &  9.92 &  9.48 &  4.4 & 100 & 6.61 & 175 & 0.071 \\
ID15 & 69.399 & 100.69 & 53.39 & 47.0 & 100 & 64.96 & 92.04 & 2.729 \\
ID16 & 93.265 & 15.14 & 14.85 &  1.9 & 2.97 & 33.88 & 105 & 5.138 \\
ID17 & 95.664 &  7.07 &  7.07 &  0.0 & 1.0 & 17.37 & 89.24 & 0.017 \\
ID18 & 92.524 & 24.80 & 24.66 &  0.6 & 118 & 19.37 & 94.64 & 8.644 \\
ID19 & 91.864 & 57.36 & 56.51 &  1.5 & 101 & 16.58 & 106 & 0.013 \\
ID20 & 91.116 & 21.64 & 21.98 & -1.6 & 104 & 12.61 & 100 & 7.471 \\
ID21 & 93.226 & 11.26 & 11.26 &  0.0 & 100 & 13.42 & 98.75 & 8.191 \\
\hline
\textbf{Overall} & \textbf{88.77} & \textbf{24.72} & \textbf{21.42} & \textbf{13.35} &
\textbf{88.72} & \textbf{20.20} & \textbf{74.64} & \textbf{0.209} \\
\hline
\end{tabular}%
}
\label{tab:multitask_results}
\end{table*}

\section{Evaluation and Analysis of Ground-Truth Data}
This section presents a comprehensive evaluation of the ground-truth annotations and the corresponding system outputs across all speech processing sub-tasks. The organizers released a mock dataset. The test data sets were released on three different occasions. The present analysis is based on mock data for which the ground-truth was released. However, the results of the remaining test data splits are submitted for evaluation by the organizers.   

\subsection{Analysis of SAD}
The speech activity detection (SAD) analysis compares the ground-truth annotations with the predictions from the pre-trained Silero VAD~\cite{Silero-VAD} model on the mock dataset. The evaluation was carried out at a 10\,ms frame resolution, ensuring fine temporal alignment between the annotated and predicted speech regions. As shown in Table~\ref{tab:multitask_results}, the system achieves near-perfect frame-level correspondence with the reference, indicating highly reliable detection of speech segments. The minor false-alarm rate arises mainly from short pauses and low-energy transitions that were occasionally classified as speech. Overall, the Silero VAD demonstrates strong consistency with the ground-truth annotations, validating its effectiveness as a front-end component for downstream diarization and ASR tasks.

\begin{table}[ht]
  \centering
  \caption{Frame-level SAD performance of the Silero VAD model on the mock dataset.}
  \setlength{\tabcolsep}{3.5pt}
  \begin{tabular}{lcccccc}
    \toprule
    \textbf{Metric} & \textbf{Prec.} & \textbf{Rec.} & \textbf{F1} & \textbf{Miss} & \textbf{FA} & \textbf{Acc.} \\
    \midrule
    Value & 0.9956 & 0.9946 & 0.9951 & 0.0054 & 0.5558 & 0.9903 \\
    \bottomrule
  \end{tabular}
  \label{tab:sad_performance}
\end{table}

\subsection{Analysis of SD}
The speaker diarization results on the annotated data distributed for preliminary analysis are shown in Table~\ref{tab:multitask_results}.

\subsection{Analysis of SID}

The speaker identification (SID) analysis was conducted using the cosine-similarity scores obtained from the ECAPA-TDNN embedding extractor.  
For a given enrollment recording, the reference speaker corresponded to the local label \texttt{Speaker1} in the file \texttt{ID16.ogg}.  
Segments from this same speaker formed the \emph{genuine} score distribution, while all remaining segments across the dataset were treated as \emph{impostor} scores.

It is important to note that \texttt{ID16.ogg} served as the sole enrollment recording in this analysis. Therefore, the SID scores reported in Table~\ref{tab:multitask_results} are meaningful only for \texttt{ID16}, whereas the values for other recordings (which do not contain the enrolled speaker) appear saturated. The per-file entries in the table thus represent relative matching indices rather than true error rates. The analytical computation of the Identification Error Rate (IER) presented below provides the accurate system-level measure of SID performance.

\begin{figure}[t]
    \centering
    \includegraphics[width=0.85\linewidth]{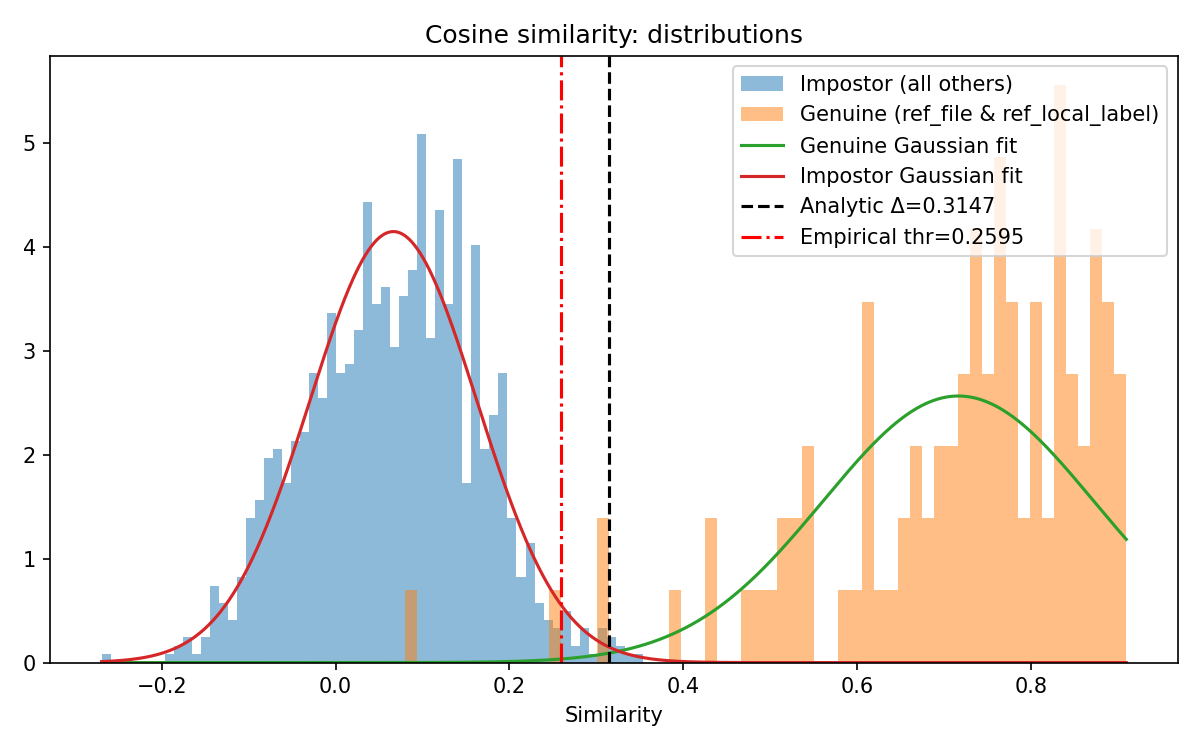}
    \caption{Empirical genuine and impostor cosine-similarity score distributions for the enrollment speaker in \texttt{ID16.ogg}.  
    The analytic threshold $\Delta = 0.3147$ lies in the low-density valley between the two distributions.}
    \label{fig:sid_scores}
\end{figure}

Figure~\ref{fig:sid_scores} illustrates the empirical distributions of genuine and impostor cosine-similarity scores.  
A clear separation is observed: impostor scores cluster primarily in the range $[0.0,0.25]$, while genuine scores occupy $[0.55,0.90]$.  
This separation yields a well-defined ``valley'' region in the interval $[0.25,0.35]$, which is ideal for selecting an operating threshold.

To obtain a principled decision boundary, we employed the analytical threshold derived from Gaussian modeling of the two score distributions:  
\[
    \Delta = \frac{\mu_{\mathrm{g}} \sigma_{\mathrm{i}} + \mu_{\mathrm{i}} \sigma_{\mathrm{g}}}
                  {\sigma_{\mathrm{g}} + \sigma_{\mathrm{i}}}
\]
where $(\mu_{\mathrm{g}}, \sigma_{\mathrm{g}})$ and $(\mu_{\mathrm{i}}, \sigma_{\mathrm{i}})$ denote the means and standard deviations of the genuine and impostor score distributions, respectively.  
For the training data, this yields $\Delta = 0.3147$, positioned exactly in the low-density region between the two empirical distributions.

Using this threshold for binary classification (enrollment vs.\ non-enrollment), we obtain:
\begin{itemize}
    \item \textbf{True Positives (TP):} 100 segments
    \item \textbf{False Negatives (FN):} 4 segments
    \item \textbf{False Positives (FP):} 7 segments
    \item \textbf{Total reference duration:} 227.73\,s
    \item \textbf{FN duration:} 2.38\,s
    \item \textbf{FP duration:} 16.62\,s
\end{itemize}

Since SID in this setting is a binary decision task, confusion errors do not arise, and the speaker Identification Error Rate (IER) reduces to
\[
    \mathrm{IER}
    = \frac{\mathrm{FN} + \mathrm{FP}}
           {\text{total reference duration}}
    = \frac{2.38 + 16.62}{227.73}
    = 0.0834.
\]
Thus, the system misidentifies only $8.34\%$ of the enrolled speaker’s total speaking time, demonstrating strong discriminative ability of the embedding extractor and providing a robust operating point for subsequent diarization or verification tasks.

\subsection{Analysis of LID}
The results obtained on the annotated data using the fine-tuned language identification system over the pre-trained VoxLingua107~\cite{VOXLINGUA107} are are shown in Table~\ref{tab:multitask_results}.

\subsection{Analysis of ASR}

The ASR evaluation was carried out using the ground--truth transcripts
provided by the organisers (\texttt{ASR\_00.csv}) and the corresponding
ASR outputs generated by our pipeline
(\texttt{ASR\_00\_pred.csv}). Prior to computing the
word error rate (WER), we performed a consistency check of the segment
timing information. Although the majority of the segments aligned
numerically by matching the keys (\texttt{AudioFileName}, \texttt{StartTS},
\texttt{EndTS}), several segments in the ground--truth file exhibited
invalid timestamp configurations, including cases where
\texttt{EndTS}~$<$~\texttt{StartTS} and occasional missing or undefined
timestamps. These inconsistencies prevent a reliable segment-level WER
computation across the full dataset.

To obtain a robust and meaningful assessment of ASR performance without
modifying the organiser-provided annotations, two evaluation strategies were
adopted: (i) File-level WER (timestamp-independent). For each recording, all
reference transcripts were concatenated in temporal order and compared
against the corresponding concatenated ASR hypothesis. This evaluation is
unaffected by local timestamp corruption and provides a stable
corpus-level estimate. Using this method, the ASR system achieved a
micro-average WER of $\text{WER}_{\text{file}} = 0.7464$, which serves as our primary metric. The distribution of these file-level WER values across the dataset is shown in Table~\ref{tab:multitask_results}.

(ii) Segment-level WER on well-formed segments. A secondary analysis was
conducted by restricting evaluation to only those segments whose timestamps
were internally consistent, i.e., \texttt{EndTS}~$\geq$~\texttt{StartTS}.
Out of the 1300 matched segments, 1296 satisfied this criterion. On this
subset, the ASR system obtained a micro-average WER of $\text{WER}_{\text{seg, valid}} = 0.7919$. This result is consistent with the file-level score but shows slightly
higher error rates due to the increased sensitivity of short segments to
tokenisation, minor timing discrepancies, and local transcription noise.
Overall, the ASR system demonstrates stable behavior across both evaluations
methods.

\subsection{Analysis of NMT}
The evaluation of the neural machine translation (NMT) system was carried out by comparing the ground-truth English translations for the mock data with the corresponding outputs generated by our translation pipeline. The BLEU score is used for evaluating the neural machine translation system, which follows the following definition:
\[
\text{BLEU} = \text{BP} \times \exp\ \left( \frac{1}{4} \sum_{n=1}^{4} \log(p_n) \right),
\]
where the brevity penalty (BP) is given by $\text{BP} = e^{\, (1 - R/C)}$,
with $R$ = length of the reference translation, 
$C$ = length of the system translation.
Here, \(p_n\) denotes the modified \(n\)-gram precision for \(n = 1,2,3,4\).
We compute the following two forms of the BLEU (Bilingual Evaluation Understudy) score:
\begin{enumerate}
    \item Corpus-level BLEU: All reference translations from the training dataset were concatenated to form a single reference corpus, and similarly, all predicted translations were concatenated to form a single hypothesis corpus. The corpus-level BLEU score obtained was $\text{BLEU}_{\text{corpus}} = 0.2093$,with a brevity penalty of $1.0$ and the following modified $n$-gram precisions: $p_1 = 0.5958$, $p_2 = 0.2748$, $p_3 = 0.1362$ and $p_4 = 0.0860$. 
Corpus BLEU is the most meaningful estimate of the overall translation quality, as it ignores bad or mismatched timestamps and looks at the translation quality as a whole.

\item File-level BLEU: As a secondary summary, BLEU was computed independently for each audio file by concatenating all its reference translations with the predicted translations. The average file-level BLEU across the dataset was $\text{BLEU}_{\text{file}} = 0.1656$.

This score reflects the variation in translation performance across individual recordings, although it is less stable than the corpus-level metric due to a smaller amount of text available per file. 
\end{enumerate}

\section{Conclusions}
In this report, we present an integrated multilingual audio processing pipeline addressing Language Agnostic Speaker Identification and Diarization, and subsequent Transcription and Translation System. The system combines several pre-trained neural components with targeted fine-tuning and custom clustering methods to achieve reliable performance on multilingual, code-mixed, and acoustically diverse audio data. 

The overall framework encompasses six interconnected modules, namely speech activity detection, speaker diarization, speaker identification, language identification, automatic speech recognition, and neural machine translation. Each component was systematically evaluated to ensure consistency across the processing chain. 

The proposed multi-kernel spectral clustering method offered consistent gains in diarization accuracy over conventional baselines and shows the effectiveness and the robustness of the proposed diarization system. 

The speaker identification system based on ECAPA-TDNN embeddings achieved an Identification Error Rate (IER) of 8.34\%. In the language identification stage, the fine-tuned VoxLingua107 model operating on 2.0\,s overlapping windows yielded an overall Diarization Error Rate (DER) of 20.20\%, reflecting stable performance across languages and recording conditions.


 The ASR module, combining Whisper-Small for English and IndicWhisper-hi for Hindi and Punjabi, achieved a corpus-level Word Error Rate (WER) of 0.7464, while the IndicTrans2-based translation module attained a corpus-level BLEU score of 0.2093.

 Overall, the integrated pipeline demonstrates stable performance across all sub-modules. Future work will focus on optimizing cross-lingual representation learning, refining overlap handling, and extending the system for real-time streaming scenarios. Further investigations will be performed on the unlabeled data once the labels become available.

\section{Acknowledgements}
The authors would like to thank the computing facilities of both institutes for providing additional computational resources.

\bibliographystyle{IEEEtran}
\bibliography{mybib}

\begin{thebibliography}{10}
\providecommand{\url}[1]{#1}
\csname url@samestyle\endcsname
\providecommand{\newblock}{\relax}
\providecommand{\bibinfo}[2]{#2}
\providecommand{\BIBentrySTDinterwordspacing}{\spaceskip=0pt\relax}
\providecommand{\BIBentryALTinterwordstretchfactor}{4}
\providecommand{\BIBentryALTinterwordspacing}{\spaceskip=\fontdimen2\font plus
\BIBentryALTinterwordstretchfactor\fontdimen3\font minus \fontdimen4\font\relax}
\providecommand{\BIBforeignlanguage}[2]{{%
\expandafter\ifx\csname l@#1\endcsname\relax
\typeout{** WARNING: IEEEtran.bst: No hyphenation pattern has been}%
\typeout{** loaded for the language `#1'. Using the pattern for}%
\typeout{** the default language instead.}%
\else
\language=\csname l@#1\endcsname
\fi
#2}}
\providecommand{\BIBdecl}{\relax}
\BIBdecl

\bibitem{SID}
R.~Togneri and D.~Pullella, ``An overview of speaker identification: Accuracy and robustness issues,'' \emph{IEEE Circuits and Systems Magazine}, vol.~11, no.~2, pp. 23--61, 2011.

\bibitem{SID-2}
R.~Jahangir, Y.~W. Teh, H.~F. Nweke, G.~Mujtaba, M.~A. Al-Garadi, and I.~Ali, ``Speaker identification through artificial intelligence techniques: A comprehensive review and research challenges,'' \emph{Expert Systems with Applications}, vol. 171, p. 114591, 2021.

\bibitem{PARK2022101317}
T.~J. Park, N.~Kanda, D.~Dimitriadis, K.~J. Han, S.~Watanabe, and S.~Narayanan, ``A review of speaker diarization: Recent advances with deep learning,'' \emph{Computer Speech \& Language}, vol.~72, p. 101317, 2022.

\bibitem{language_recognition}
S.~Dey, M.~Sahidullah, and G.~Saha, ``An overview of indian spoken language recognition from machine learning perspective,'' \emph{ACM Trans. Asian Low-Resour. Lang. Inf. Process.}, vol.~21, no.~6, 2022.

\bibitem{dey2024towards}
------, ``Towards cross-corpora generalization for low-resource spoken language identification,'' \emph{IEEE/ACM Transactions on Audio, Speech, and Language Processing}, 2024.

\bibitem{ASR}
S.~Alharbi, M.~Alrazgan, A.~Alrashed, T.~Alnomasi, R.~Almojel, R.~Alharbi, S.~Alharbi, S.~Alturki, F.~Alshehri, and M.~Almojil, ``Automatic speech recognition: Systematic literature review,'' \emph{IEEE Access}, vol.~9, pp. 131\,858--131\,876, 2021.

\bibitem{stahlberg2020neural}
F.~Stahlberg, ``Neural machine translation: A review,'' \emph{Journal of Artificial Intelligence Research}, vol.~69, pp. 343--418, 2020.

\bibitem{Silero-VAD}
S.~Team, ``Silero vad: pre-trained enterprise-grade voice activity detector (vad), number detector and language classifier,'' \url{https://github.com/snakers4/silero-vad}, 2024.

\bibitem{Raghav2026MKSGCSC}
N.~Raghav, A.~Gupta, S.~Das, and M.~Sahidullah, ``{MK-SGC-SC}: Multiple kernel guided sparse graph construction in spectral clustering for unsupervised speaker diarization,'' 2026, {(Submitted to ICASSP 2026)}.

\bibitem{speechbrain_v1}
M.~Ravanelli \emph{et~al.}, ``Open-source conversational {AI} with {SpeechBrain} 1.0,'' \emph{Journal of Machine Learning Research}, vol.~25, no. 333, 2024.

\bibitem{speechbrain}
M.~Ravanelli, T.~Parcollet, P.~Plantinga, A.~Rouhe, S.~Cornell, L.~Lugosch, C.~Subakan, N.~Dawalatabad, A.~Heba, J.~Zhong, J.-C. Chou, S.-L. Yeh, S.-W. Fu, C.-F. Liao, E.~Rastorgueva, F.~Grondin, W.~Aris, H.~Na, Y.~Gao, R.~D. Mori, and Y.~Bengio, ``{SpeechBrain}: A general-purpose speech toolkit,'' 2021, arXiv:2106.04624.

\bibitem{desplanques20_interspeech}
B.~Desplanques, J.~Thienpondt, and K.~Demuynck, ``{ECAPA-TDNN}: Emphasized channel attention, propagation and aggregation in tdnn based speaker verification,'' in \emph{Proc. INTERSPEECH}, 2021, pp. 3830--3834.

\bibitem{SC-pNA}
N.~Raghav, A.~Gupta, M.~Sahidullah, and S.~Das, ``Self-tuning spectral clustering for speaker diarization,'' in \emph{Proc. ICASSP}, 2025, pp. 1--5.

\bibitem{VOXLINGUA107}
J.~Valk and T.~Alumäe, ``Voxlingua107: A dataset for spoken language recognition,'' in \emph{Proc. IEEE Spoken Language Technology Workshop (SLT)}, 2021.

\end{thebibliography}

\end{document}